\documentclass[%
 preprint,%
 amssymb, amsmath,%
 aip,cha,%
]{revtex4-1}
\usepackage{graphicx, setspace}
\usepackage{docs}%
\usepackage{bm}%
\usepackage[colorlinks=true,linkcolor=blue]{hyperref}
\expandafter\ifx\csname package@font\endcsname\relax\else
\expandafter\expandafter
\expandafter\usepackage
\expandafter\expandafter
\expandafter{\csname package@font\endcsname}%
\fi
\begin{document}

\title{Nonlinear Dynamics of Relativistically Intense Cylindrical and Spherical Plasma Waves}%

\author{Arghya Mukherjee}%
\email{arghya@ipr.res.in}
\author{Sudip Sengupta}
\affiliation{Institute for Plasma Research, Gandhinagar, India, 382428}
\affiliation{Homi Bhabha National Institute, Training School Complex, Mumbai, India 400094}

%\date{March 2010}%
%\revised{August 2010}%

\begin{abstract}
Spatio-temporal evolution and breaking of relativistically intense cylindrical and spherical space charge oscillations in a homogeneous cold plasma is studied analytically and numerically using Dawson Sheet Model [J.M. Dawson, Phys. Rev.{\bf{113}}, 383 (1959)]. It is found that cylindrical and spherical space charge oscillations break via the process of phase mixing at an arbitrarily small amplitude due to anharmonicity introduced by geometry and relativistic mass variation effects. A general expression for phase mixing time (wave breaking time) has been derived and it is shown that for both cases, it scales inversely with the cube of the initial wave amplitude. Finally this analytically obtained scaling is verified by using a numerical code based on Dawson Sheet Model.
\end{abstract}

%\pacs{52.35.Mw, 52.27.Ny, 52.65.Rr}

\maketitle

%\tableofcontents

\section{Introduction}
The space-time evolution of nonlinear plasma oscillations/waves and their breaking is a fascinating field of study due to its far reaching applications in nonlinear plasma physics \cite{a, b, 7}. The breaking of relativistically intense plasma waves generated by a high intensity laser/particle beam pulse plays a crucial role in particle acceleration processes \cite{e, f, g, i, j, c} and inertial confinement fusion experiments \cite{a1, a2, ssnf}. Dawson \cite{1} was the first to elucidate the fact that for a nonrelativistic cold plasma, oscillations in slab geometry would be stable below a critical amplitude, called wave breaking limit which is given by $eE/m\omega _pv_\phi \approx 1$ ($e$ and $m$ respectively being the charge and mass of an electron; $\omega _p$ is the nonrelativistic plasma frequency, $v_\phi$ is the phase velocity of the plasma wave and $E$ is the self-consistent electric field amplitude). Beyond this limit multistream flow or fine scale mixing develops which destroys the collective motion of the plasma electrons and hence the oscillations break within a plasma period.

Even without approaching the above ``wave breaking limit'', Dawson pictured a novel
phenomenon where plasma oscillations start losing it's periodicity gradually with
time, provided that the characteristic frequency of oscillation, due to some physical reasons, acquires a spatial dependency. In this case, oscillations break at a particular time decided by the initial amplitude which is far below the corresponding wave breaking limit. This phenomenon is called ``phase mixing'' \cite{4,5,6,8,k,16, 3, arsu}. Due to spatial dependency of the characteristic frequency, neighbouring electrons gradually move out of phase and eventually cross each other causing the wave to break at arbitrarily small initial amplitude \cite{4, 5, 6, 8, k, 16, 3, arsu}. Thus due to phase mixing, oscillations/waves break at arbitrarily small initial amplitude far below the corresponding ``wave breaking limit''. Dawson derived a general expression for phase mixing time scale (wave breaking time) by using a physical reasoning which is based on out of phase motion of neighbouring oscillators separated by a distance equal to twice the amplitude of the oscillation/wave, and demonstrated that for ``nonrelativistic oscillations'' phase mixing can occur if there is a density inhomogeneity (either fixed \cite{ref244} or self-generated \cite{ref233}) or the geometry of the oscillation changes from planar to cylindrical/spherical \cite{1}.

%Dawson further extended his nonrelativistic calculations by including geometrical effects where the electrons are oscillating back and forth along the radii of either a cylinder or sphere and demonstrated that in the nonrelativistic case, cylindrical or spherical oscillations/waves always break at an arbitrary amplitude of the applied perturbation. Physically it occurs via a well known process called phase mixing \cite{4,5,6,8,k,16}. This phase mixing is caused by temporal dependence of the phase difference between oscillating electrons constituting the oscillation/wave which gradually results in crossing of neighbouring electron orbits resulting in singularity in the electron density. 
Later it has been shown by several authors that plasma oscillations/waves in a slab geometry can also break in the similar fashion, if the electron's quiver velocity becomes relativistic \cite{drake, 3, 4, 5, 6, 8, k, 16, 3, arsu}. In this case phase mixing time scale crucially depends on the amplitude of the initialised oscillation/wave. By using Dawson Sheet Model, the authors in Refs. \cite{4, 5} derived an analytical expression for phase mixing time scale of a relativistically intense oscillation/wave in a slab geometry as a function of the initial amplitude. These authors \cite{4, 5} also confirmed their theoretical predictions by using a code based on Dawson Sheet Model and observed that the phase phase time scale of a relativistically intense oscillation/wave in a slab geometry is inversely proportional to the cube of the initial amplitude.
%In last few years some intensive attention has been taken in order to study cylindrical and spherical plasma oscillations/waves with relativistic effects both theoretically\cite{17, 18, 19, 20} and experimentally\cite{21, 22} by considering a tightly focused laser pulse in an underdense plasma. By using Lagrange coordinates, Gorbunov \textit{et. al.}\cite{17,18} and Bulanov \textit{et. al.}\cite{19} respectively studied the evolution of cylindrical and spherical plasma waves. These authors respectively derived the equation of motion of an electron oscillating along the radii of a cylinder and sphere. The frequency of oscillation correct upto second order in respective cases are also derived, surprisingly which are appeared to be same. But, an analytical expression for phase mixing time as a function of the amplitude of the applied perturbation for cylindrical and spherical plasma oscillations/waves are yet to be presented as a function of the amplitude of the applied perturbation as done by Sengupta \textit{et. al.}\cite{4,5} for relativistic planar plasma oscillations/waves. Though Gorbunov \textit{et. al.} \cite{17, 18} attempted to predict a phase mixing time by using Dawson's argument \cite{1}, still the verification of their prediction was never been shown explicitly. Therefore, the investigation of variation of phase mixing time scale of relativistically intense plasma oscillations/waves with the amplitude of the applied perturbation in a cylindrically and spherically symmetric system forms a concrete area of research, which is explicitly explored in this paper.

In the last few years much attention has been paid to the study of relativistically intense cylindrical and spherical plasma oscillations/waves, both theoretically \cite{17, 18, 19, 20} and experimentally \cite{21, 22}. For example, Gorbunov et. al. \cite{17,18} and Bulanov et. al. \cite{19} respectively studied the evolution of cylindrical and spherical plasma waves analytically by using Lagrange coordinates \cite{a}. These authors respectively derived the equation of motion of an electron oscillating along the radius of a cylinder \cite{17, 18} and sphere \cite{19}. The frequency of oscillation correct upto second order in oscillation amplitude were derived for cylindrical and spherical case, which turned out to be same. In the former case \cite{17, 18}, the authors observed trajectory crossing of the neighbouring electrons which leads to wave breaking via phase mixing. In the latter case \cite{19}, the authors observed that after some plasma period, the wave changes it's direction of propagation which also occurs due to spatial dependency of the characteristic frequency of the wave. This time was termed as
``turn-around time'' \cite{19}. But, an analytical expression for phase mixing time as a
function of the amplitude of the applied perturbation for cylindrical and spherical
plasma oscillations/waves were not presented. Though Gorbunov et. al. \cite{17, 18} attempted to predict a phase mixing time by using Dawson's argument \cite{1}, still the verification of their prediction was never shown explicitly. %Therefore the space time evolution of relativistically intense cylindrical/spherical plasma oscillations/waves, estimation of phase mixing time and their variation with the amplitude of the applied perturbation has been explicitly explored in this paper.

In this paper, we extend Dawson's earlier work on cylindrical and spherical oscillations by including relativistic mass variation effect of the electrons. 
%An important feature of cylindrical and spherical oscillations is that it's parameters vary only with the distance from the source i.e they depend only on the radial coordinate of the oscillating electrons. Therefore we limit our study to one dimensional case only. 
In section -\ref{sec:2} we extend Dawson Sheet Model from planar to cylindrical and spherical geometry and also include the relativistic mass variation effects. We first derive the expressions for the fluid variables viz. density ($n$), electric field ($E$) and velocity ($v$) by respectively using the principle of conservation of number of particles (continuity equation), Gauss's Law and the relativistic equation of motion. Then we solve the equation of motion in respective coordinate system by using Lindstedt-Poincar\'{e} perturbation method \cite{23} and derive an approximate expression for frequency of oscillation in the weakly relativistic limit upto fourth order in oscillation amplitude. We observe that in general the expressions for frequencies acquire spatial dependency which ultimately lead to breaking via phase mixing. In section -\ref{sec:3} we study the dynamics of cylindrical and spherical plasma oscillation and derive analytical expressions for (cylindrical and spherical) phase mixing time scales as a function of initial amplitude by using Dawson's argument \cite{1}. Further, we verify this scaling by performing numerical simulations, using a code based on Dawson Sheet Model \cite{1, 4, 5, 6, 8, 16, 25, arsu} by extending it to cylindrical and spherical geometry. Finally in section -\ref{sec:4} we summarize this work.
%study the space-time evolution and breaking of nonlinear oscillations of plasma electrons in a cylindrically and spherically symmetric system by using Dawson Sheet Model in both relativistic and nonrelativistic limit. An important feature of such nonlinear cylindrical and spherical oscillations is that its parameters vary with distance from the source i.e it depend only on the radial coordinate of the oscillating electron. Therefore we limit our study in one dimensional case only. In section \ref{sec:2} we first derive the expressions for the fluid variables \textit{viz.} $n$, $E$ and $v$ by respectively using the principle of conservation of number of particles, Gauss's Law and Newton's Law which is found to be an easier way to derive the fluid variables rather than to handle the Lagrange coordinates\cite{a,k,24}. Then we solve the equation of motion by using Lindstedt-Poincare method \cite{23} correct upto third order correction considering small amplitude $\&$ weakly relativistic limit. The frequency of oscillation is also given in this section for both two cases. In section \ref{sec:3} we study the dynamics of cylindrical and spherical plasma oscillation and present a scaling law for phase mixing time by using Dawson's argument\cite{1} as a function of the amplitude of the applied perturbation for all the cases under consideration. Further, we verify this scaling numerically, using a code based on Dawson Sheet Model\cite{1,4,5,6,8,16,25,arsu}. Finally section in \ref{sec:4} we summarize our results. 

\section{Governing Equations} \label{sec:2}
According to Dawson Sheet Model \cite{4, 5, 6, 8, 16, 25, arsu}, cylindrical and spherical oscillations arise respectively due to cylindrical and spherical sheet of charges oscillating about their equilibrium positions. These sheet of charges are embedded in a homogeneous background of immobile ions. Unlike the slab geometry, in these cases the electric field becomes space dependent due to geometrical effects {\it i.e.} for same displacement amplitude the electric field amplitude becomes different for different equilibrium positions of the sheets, which in turn makes the frequency of oscillation space dependent. This is the basic reason why phase mixing and wave breaking in these cases is an inherent phenomena, arising due to geometry of the problem \cite{1, 17, 18, 19, 20}.

%In this section we take Dawson's original work \cite{1} and introduce relativistic mass variation effects and derive expressions for the fluid variables \textit{viz.} $n$, $v$ and $E$.
% can be derived in a more convenient way. 
%However, the fluid variables can also be obtained from the fluid equations using Lagrange Coordinates\cite{a, k, 24} which has been taken up by several authors \cite{17, 18, 19, 20}.
In the following subsections we respectively derive the expressions for the fluid variables and derive equation of motion for cylindrical $\&$ spherical oscillations considering relativistic mass variation effect of the electrons. Let $r_0$ and $R(r_0,t)$ respectively be the equilibrium positions and displacement from the equilibrium positions of the electron sheets. So the Euler positions of the sheets can be written as $r(r_0,t) = r_0 + R(r_0,t)$.      

\subsection{Fluid Variables for Cylindrical Oscillations} 
In cylindrical geometry, the conservation of the number of particle yields
\begin{equation}
2\pi n_0r_{0}dr_{0} = 2\pi nrdr \nonumber \\
\end{equation}
where, $n_0$ and $n$ respectively are the equilibrium and instantaneous number density. Using the expression for $r(r_{0},t)$ gives number density as
\begin{equation}
n(r_0,t) = \frac{n_0r_0}{(r_0 + R)(1+\frac{\partial R}{\partial r_0})} \label{eq:1}
\end{equation}
Now using Gauss's Law in cylindrical geometry, we have
\begin{equation}
\frac{1}{r}\frac{\partial}{\partial r}(rE) = 4\pi e(n_0 - n) \nonumber\\
\end{equation}
Substituting the value of $n$, from Eq.(\ref{eq:1}), the electric field stands as
\begin{equation}
E(r_0,t) = 2\pi en_0\left[\frac{(r_0 + R)^2 - r_0^2}{(r_0 + R)}\right] \label{eq:2}
\end{equation}
 Now the relativistically correct equation of motion of electron sheet can be written as
\begin{equation}
m\frac{d}{dt}\left[\frac{\dot{R}}{(1-\frac{\dot{R}^2}{c^2})^{1/2}}\right] = -eE \nonumber \\
\end{equation}
putting the value of $E$ from Eq.(\ref{eq:2}), we get
\begin{equation}
\frac{\ddot{R}}{(1-\frac{\dot{R}^2}{c^2})^{3/2}} = -\frac{\omega_p^2}{2}\left[\frac{(r_0 + R)^2 - r_0^2}{(r_0 + R)}\right]   \label{eq:3}
\end{equation}
Here $\omega_p = 4\pi n_0e^2/m$ is the nonrelativistic plasma frequency. Here $dot$ sign represents derivative w.r.t time.
Taking $R/r_0 = \rho$, Eq.(\ref{eq:3}) modifies as
\begin{equation}
\frac{\ddot{\rho}}{(1 - \frac{r_0^2\dot{\rho}^2}{c^2})^{3/2}} + \frac{\omega_p^2}{2}\left[\frac{(1 + \rho)^2 - 1}{(1+\rho)}\right] = 0 \label{eq:4}
\end{equation}
In nonrelativistic limit ($c \rightarrow \infty$), we get the same equation as obtained by Dawson \cite{1}
\begin{equation}
\ddot{\rho} + \frac{\omega_p^2}{2}\left[\frac{(1 + \rho)^2 - 1}{(1+\rho)}\right] = 0 \nonumber
\end{equation}
 In the weakly relativistic limit and small amplitude oscillations Eq.(\ref{eq:4}) can be simplified as \cite{17, 18} (expanded upto the third order of $\rho$ and second order of $\dot{\rho}$) 
\begin{equation}
\ddot{\rho} - \frac{3}{2}\frac{r_0^2\omega_p ^2}{c^2}\rho\dot{\rho}^2 + \omega_p^2 \rho - \frac{\omega_p^2}{2}\rho ^2 + \frac{\omega_p^2}{2}\rho ^3 = 0   \label{eq:5}
\end{equation}
Now, by using Lindstedt - Poincar\'{e} perturbation method \cite{23}, the expression for frequency correct upto the fourth order of oscillation amplitude can be written as 
\begin{equation}
\Omega_{cy}(rel)  = \omega_p\left[1 + \frac{\rho_0(r_0)^2}{12} + \frac{\rho_0(r_0)^4}{512} - \frac{3\omega_p^2}{16}\frac{r_0^2\rho_0(r_0)^2}{c^2} \right] \label{eq:6}
\end{equation}
Here $\rho_0(r_0)$ is the displacement amplitude of the electrons, which in general depends on their equilibrium positions $r_0$. In nonrelativistic limit $c\rightarrow \infty$, Eqs.(\ref{eq:5}) and (\ref{eq:6}) respectively become
\begin{equation}
\ddot{\rho} + \frac{\omega_p^2}{2}\left[2\rho - \rho ^2 + \rho ^3\right] = 0  \label{eq:7}
\end{equation}
and
\begin{equation}
\Omega_{cy}(nonrel)  = \omega_p\left[1 + \frac{\rho_0(r_0)^2}{12} + \frac{\rho_0(r_0)^4}{512}\right]    \label{eq:8}
\end{equation}

 \subsection{Fluid Variables for Spherical Oscillations}
 Now following the same procedure in spherical geometry the fluid variables can be written as
\begin{equation}
 n(r_0,t) = \frac{n_0r_0^2}{(r_0 + R)^2(1 + \frac{\partial R}{\partial r_0})} \label{eq:9}
\end{equation}   
%\begin{equation}
%\frac{1}{r^2}\frac{\partial (r^2E)}{\partial r} = 4\pi e(n_0-n)  \nonumber \\
%\end{equation}
\begin{equation}
E(r_0,t) = \frac{4\pi en_0}{3}\left[\frac{(r_0 + R)^3 - r_0^3}{(r_0 + R)^2} \right] \label{eq:10}
\end{equation}
Using the above expression or electric field, the relativistically correct equation of motion of an electron sheet oscillating along the radius of a sphere is given by,
\begin{equation}
 \frac{\ddot{R}}{(1-\frac{\dot{R}^2}{c^2})^{3/2}} + \frac{\omega_p^2}{3}\left[\frac{(r_0 + R)^3 - r_0^3}{(r_0 + R)^2}\right] = 0  \label{eq:11}
 \end{equation}
 In nonrelativistic limit, the above Eq. transforms to \cite{1}
 \begin{equation}
\ddot{R} + \frac{\omega_p^2}{3}\left[\frac{(r_0 + R)^3 - r_0^3}{(r_0 + R)^2}\right] = 0  \nonumber
 \end{equation}
 In terms of $\rho$, Eq.(\ref{eq:11}) becomes
 \begin{equation}
 \frac{\ddot{\rho}}{(1 - \frac{r_0^2\dot{\rho}^2}{c^2})^{3/2}} + \frac{\omega_p^2}{3}\left[\frac{(1 + \rho)^3 - 1}{(1+\rho)^2}\right] = 0 \label{eq:12}
 \end{equation}
 In weakly relativistic limit and small amplitude oscillation, the above Eq. takes the form \cite{19}
 \begin{equation}
\ddot{\rho} - \frac{3}{2}\frac{r_0^2\omega_p ^2}{c^2}\rho \dot{\rho}^2 + \omega_p^2 \rho - \omega_p^2 \rho ^2 + \frac{4\omega_p^2}{3}\rho ^3 = 0  \label{eq:13}
\end{equation}
 And frequency of oscillation stands as (using Lindstedt - Poincar\'{e} perturbation method \cite{23})
 \begin{equation}
 \Omega_{sph}(rel) = \omega_p\left[1 + \frac{\rho_0(r_0)^2}{12} + \frac{\rho_0(r_0)^4}{72}- \frac{3\omega_p^2}{16}\frac{r_0^2\rho_0(r_0)^2}{c^2} \right] \label{eq:14} 
 \end{equation}
 In non-relativistic limit, the equation of motion for small amplitude oscillation becomes \cite{1}
 \begin{equation}
\ddot{\rho} + \frac{\omega_p^2}{3}\left[3\rho - 3\rho ^2 + 4\rho ^3\right] = 0  \label{eq:15}
\end{equation}
And the frequency of oscillation can be written as
\begin{equation}
 \Omega_{sph}(nonrel) = \omega_p\left[1 + \frac{\rho_0(r_0)^2}{12} + \frac{\rho_0(r_0)^4}{72}\right] \label{eq:16} 
 \end{equation}
 Here we want to note that Eqs.(\ref{eq:5}) and (\ref{eq:13}) respectively are the same equation derived by Gorbunov \cite{17, 18} and Bulanov \cite{19}. In the expressions of $\Omega_{cy}(rel)$ and $\Omega_{sph}(rel)$ the second term represents correction due to the relativistic effects and other terms represent additional anharmonicity introduced by cylindrical and spherical geometry respectively. In addition to this, it should be noted from the expressions of the electric field [Eqs.(\ref{eq:2}) and (\ref{eq:10})] that, even for the same displacement amplitude, the electric field depends explicitly on the equilibrium positions of the electrons (unlike the planar geometry case) which results in a position dependent restoring force. Therefore the frequency of oscillation depends on the equilibrium positions of the sheets which leads to wave breaking via phase mixing. In the next section we present a scaling law for this phase mixing time.

 \section{Dynamics of Cylindrical $\&$ Spherical plasma oscillations and Calculation of Phase Mixing Time} \label{sec:3}

In this section we describe the dynamics of relativistically intense cylindrical and spherical plasma oscillations and calculate their phase mixing time. In order to study the space-time evolution and breaking of cylindrical and spherical oscillations, we first load the initial conditions needed to excite axisymmetric cylindrical and spherical oscillations respectively in an one dimensional (along the radius) sheet code (based on Dawson Sheet Model) containing $\sim$ 10000 cylindrical and spherical surfaces of charges. For cylindrical and spherical oscillations, its parameters depend only on the radial coordinate of the oscillating species (here the electron sheets) \textit{i.e.} they are azimuthally symmetric in nature. As Bessel functions and Spherical Bessel functions \cite{14,15} respectively form a complete orthogonal set (basis functions) in cylindrical and spherical coordinate systems, then any arbitrary radial perturbation imposed in these systems can be written as a superposition of these basis functions (Fourier Bessel Series) in their respective coordinate system \cite{14,15}. Therefore to excite an oscillation in cylindrically and spherically symmetric system we respectively use Bessel functions and Spherical Bessel functions as an initial perturbation. Here, we note that, the structure and propagation of nonrelativistic axisymmetrical waves in linear and nonlinear regime using such type of initial conditions has been studied first by Travelpiece $\&$ Gould \cite{10} for a cylindrical plasma column and later continued by several authors \cite{11, 12, 13}. Using these type of initial conditions the equation of motion for each electron sheet is solved using fourth order Runge-Kutta scheme. At each time step, ordering of the sheets is checked for sheet crossing. Phase mixing time is measured as the time taken by any two of the adjacent sheets to cross over \cite{1, 4, 5, 6, 8, k, 16, 17, 18, 22, arsu}. 

In the following subsections we study the space-time evolution $\&$ derive the expressions for phase mixing time as a function of the amplitude of the applied perturbation for cylindrical and spherical plasma oscillations respectively.

\subsection{Phase Mixing of Relativistically Intense Cylindrical Plasma Oscillations}
The relativistic equation of motion of an electron sheet along the radius of a cylinder is given by Eq.(\ref{eq:3}). This equation can be solved numerically with the help of two initial conditions $R(r_0,t = 0)$ and $\dot{R}(r_0,t = 0)$. Here we consider plasma oscillations localized in space in the vicinity of the axis $r = 0$ \cite{17, 18}. We assume that electron velocity at $t = 0$ is zero : $\dot{R}(r_0,t = 0) = 0$. We also assume that at $t = 0$ the oscillations are excited by an electric field of the form \cite{10, 11, 12}
\begin{equation}
\bar{E}(r_0,t = 0) = \frac{eE(r_0,t = 0)}{m\omega_p v_\phi} = \Delta J_n\left[\frac{\alpha _{n m} r(r_0, 0)}{R_0}\right]    \label{eq:17}
\end{equation}
 For relativistically intense oscillations, $v_\phi \rightarrow c$. $J_n$ is $n$-th order Bessel function. $\alpha_{n m}$ is the $m$-th zero of $n$-th order Bessel function \cite{14, 15}. $\Delta$ is the amplitude of applied electric field perturbation and $R_0$ is the maximum value of the radius of the cylindrical system under consideration. We consider that at initial time, the electric field $E(r_0,0)$ at the boundaries of the simulation are zero i.e $E(r_0 = 0, t = 0) = 0$ and $E(r_0 = R_0, t = 0) = 0$. 
 
In the above scenario all initial conditions are satisfied by the lowest order mode is $n = 1$ and $m = 1$ (As E is zero on the axis at $t = 0$ so $n = 0$ cannot be taken as lowest order mode as $J_0(\alpha_{01}r/R_0) \neq 0$ at $r = 0$). Therefore, here we study the space-time evolution of the lowest order mode $\Delta J_1(\alpha_{11}r/R_0)$. The length of the system $R_0$ is taken upto the first zero of the Bessel function.

To solve Eq.(\ref{eq:3}) we first numerically calculate the initial profile of $R(r_0, t = 0)$ in the following way: At initial instant of time $t = 0$, sheets are at their equilibrium position $r_0$ and radially displaced from their equilibrium positions by an amount $R(r_0, 0)$ such that they produce an electric field  perturbation given by Eq.(\ref{eq:17}) {\it i.e.} $\bar{E}(r_0,t = 0) = \Delta J_1[\alpha _{11} r(r_0,0)/R_0]$.  ( The Euler positions of the electrons at $t = 0$ become $r(r_0, 0) = r_0 + R(r_0, 0)$. )
%All the particles at their initial position produce an electric field of the form
%$\bar{E}(r_0,t = 0) = \Delta J_1[\alpha _{11} r(r_0,0)/R_0]$. 
On the other hand from Gauss's Law the electric field $E(r_0, 0)$ is given by 
\begin{equation}
E(r_0, 0) = 2\pi en_0\frac{[r_0 + R(r_0, 0)]^2 - r_0^2}{[r_0 + R(r_0, 0)]}  \label{eq:18}
\end{equation}
Comparing Eqs.(\ref{eq:17}) and (\ref{eq:18}) we find $R(r_0, 0)$ for given values of $r_0$. 

\begin{figure}[htbp]
\center
\includegraphics[width=0.6\textwidth]{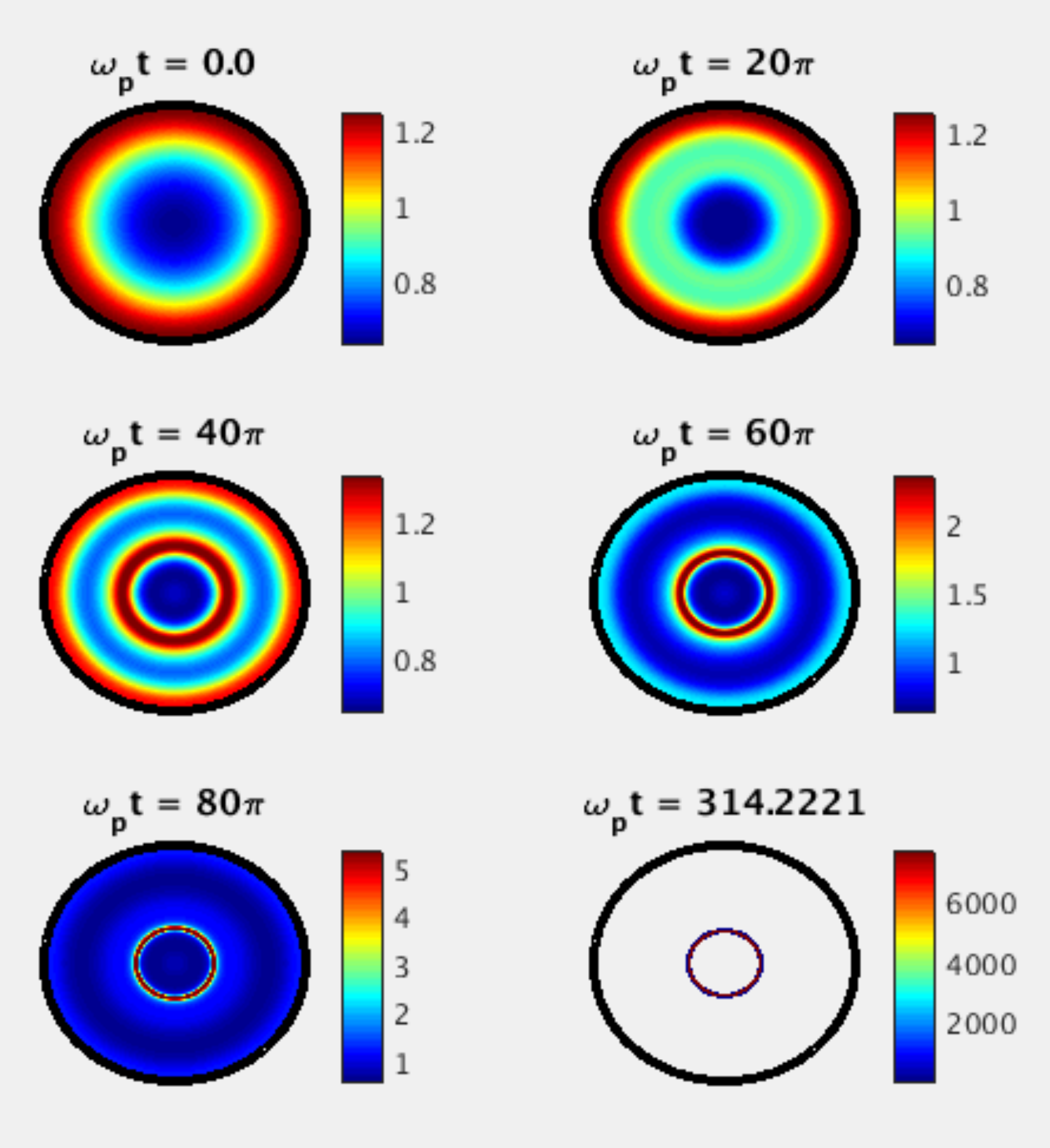}  
\caption{(Color online) Time evolution of density of relativistic cylindrical oscillation for amplitude $\Delta = 0.5$.}  
\label{fig:fig1} 
\end{figure}

\begin{figure}[htbp]
\center
\includegraphics[width=0.6\textwidth]{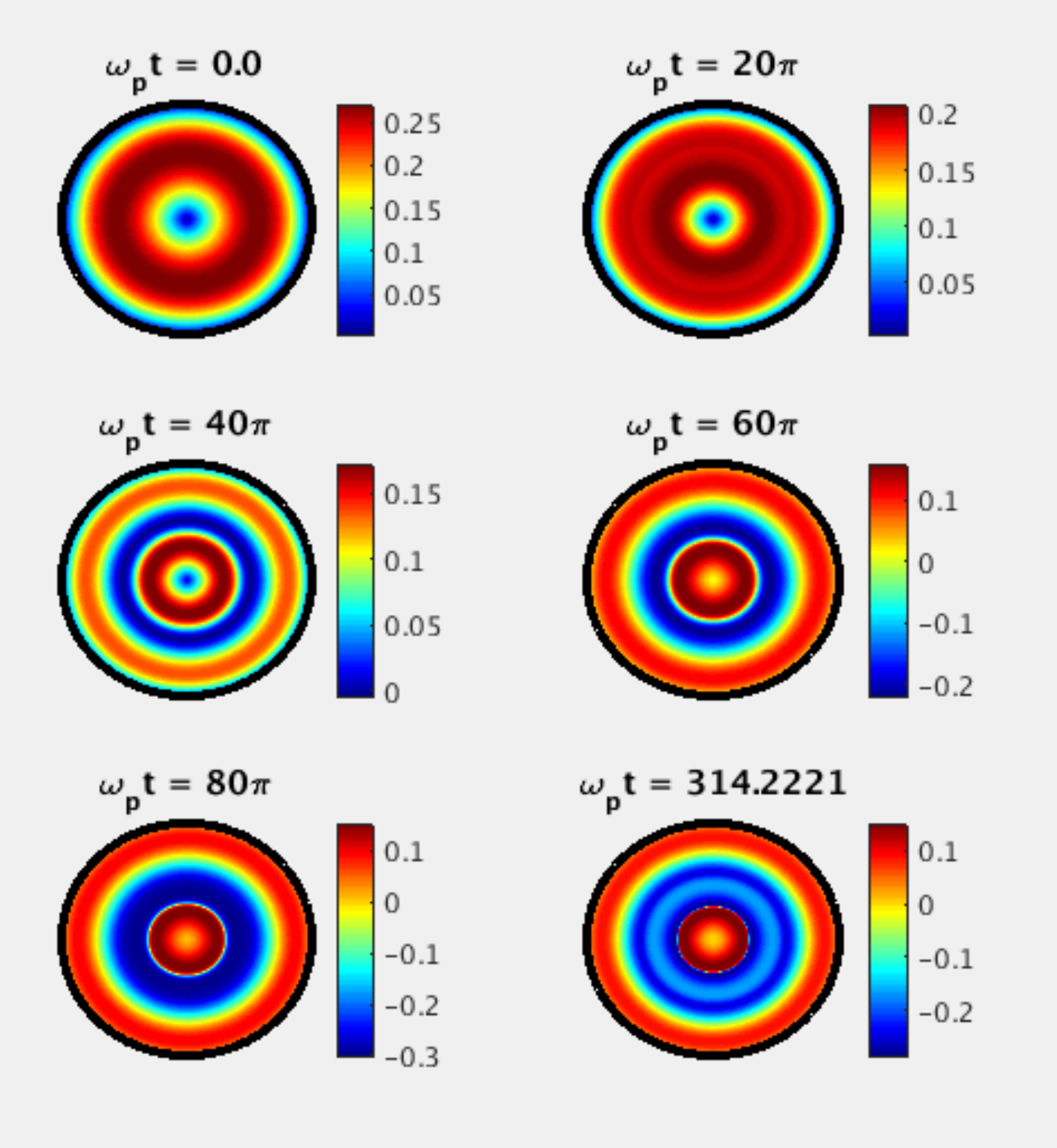}  
\caption{(Color online) Time evolution of electric field of relativistic cylindrical oscillation for amplitude $\Delta = 0.5$.}  
\label{fig:fig2} 
\end{figure}

Using the above initial conditions, numerical computations have been carried out and the spatial variation of electric field and electron density with time have been shown. The results of relevant simulations are illustrated in Figs-(\ref{fig:fig1}) and (\ref{fig:fig2}). Figs-(\ref{fig:fig1}) and (\ref{fig:fig2}) respectively show the show the snapshots of the electron density and electric field profiles for the value $\Delta = 0.5$. Fig-(\ref{fig:fig1}) shows that, as time progresses, the density maxima increases gradually and shows a high spike at $\omega _pt = 314.2221$ which is a signature of wave breaking via the process of phase mixing \cite{1, 4, 5, 6, 8, k, 16, 17, 18, 22, arsu}. From Fig-(\ref{fig:fig2}) we observe that, as time goes on, the radial profile of electric field becomes steeper and acquires a jump at the off-axis radial point, where electron density spikes.

Now we calculate the phase mixing time scale in the following way: Equating the electric field given by Eq.(\ref{eq:17}) with Eq.(\ref{eq:18}) and in the small amplitude limit $\Delta << 1$, we can write  
\begin{equation}
\frac{\omega _pr_0}{v_\phi}\rho _0(r_0) \sim \Delta J_1 \left( \alpha _{11}\frac{r_0}{R_0}\right)      \label{eq:19}
\end{equation}
Substituting the value of $\rho_0(r_0)$ from above in the expression for frequency of relativistic cylindrical plasma oscillation [Eq.(\ref{eq:6})] we get (correct upto $\Delta ^2$)
\begin{equation}
\Omega_{cy}(rel)  \sim \omega_p\left[1  - \frac{3}{16}\frac{v_\phi ^2\Delta ^2}{c^2}J_1 ^2\left( \alpha _{11}\frac{r_0}{R_0}\right) + \frac{v_\phi ^2\Delta ^2}{12\omega _p^2r_0^2}J_1 ^2\left( \alpha _{11}\frac{r_0}{R_0}\right) \right] \label{eq:20}
\end{equation}
According to Dawson's argument \cite{1}, the phase mixing time $(\omega_p \tau_{mix})$ depends on the spatial derivative of frequency as $\omega_p \tau _{mix} \sim \pi/[2R_{max}(d\Omega/dr_{0})]$. Differentiating Eq.(\ref{eq:20}) w.r.t $r_0$ and noting $R_{max}$ is proportional to $\Delta$, the phase mixing time scale can be written as
\begin{equation}
\omega _p\tau_{mix} \sim  \frac{1}{\Delta ^3}   \label{eq:21}
\end{equation}
Here, we have omitted the proportionality constant which is a function of $r_0$. This proportionality constant is related  to the position of breaking, which generally depends on the profile of the initial perturbation and is not of general interest. In the similar manner using Eq.(\ref{eq:8}), we can find that, for nonrelativistic cylindrical oscillations also, phase mixing time scale follows the same scaling law as given by Eq.(\ref{eq:21}).

In order to verify this scaling expressed by Eq.(\ref{eq:21}) we have repeated our numerical experiment for different values of $\Delta$. The variation of phase mixing time as function of the amplitude of applied perturbation is shown in Fig-(\ref{fig:fig3}) for both relativistic and nonrelativistic oscillations. In the figure, points represent the simulation results and the solid lines represent our theoretical scaling obtained from Eq.(\ref{eq:21}). By comparing the results, we observe that for a fixed value of the applied perturbation, relativistic effect reduces the phase mixing time.

\subsection{Phase Mixing of Relativistically Intense Spherical Plasma Oscillations}
In this subsection we present the space-time evolution of spherical plasma oscillations and estimate the phase mixing time in a similar fashion as described in the above subsection.

Here the spherical oscillations have been excited by electric field $E(r_0,t)$ of the form
 \begin{equation}
\bar{E}(r_0,t = 0) = \frac{eE(r_0,t = 0)}{m\omega_p v_\phi} = \Delta j_\nu\left[\frac{\beta _{\nu m} r(r_0, 0)}{R_0}\right]    \label{eq:23}
\end{equation}
where $j_\nu(x)$ is the Spherical Bessel function of order $\nu$ and defined as $j_\nu(x) = \sqrt{\pi /2x} J_{\nu + 1/2}(x)$. $\beta_{\nu m}$ is the $m$-th zero of $\nu$-th order Spherical Bessel function \cite{14, 15}. We have computed the displacement of the sheets from their equilibrium positions {\it i.e.} the value of $R(r_0,0)$ by comparing Eq.(\ref{eq:10}) and (\ref{eq:23}) in the similar fashion as described in the previous subsection. Here we have taken the lowest order mode $\Delta j_1(\beta_{11}r/R_0)$ and the maximum radius of the system ($R_0$) is taken upto the first zero of the Spherical Bessel function. We have solved Eq.(\ref{eq:11}) numerically and calculated the density and electric field profile respectively from the expressions (\ref{eq:9}) and (\ref{eq:10}). The snapshots of density and electric field profile at different time steps are shown in Figs-(\ref{fig:fig5}) and (\ref{fig:fig6}) respectively. Like cylindrical waves here also it is observed that phase mixing leading to wave breaking is manifested by a density burst and a sharp gradient in the electric field profile.

Now to obtain a scaling law for phase mixing of spherical oscillations, we again follow the same procedure. For a small amplitude perturbation $\Delta << 1$, $\frac{\omega _pr_0}{v_\phi}\rho _0(r_0) \sim \Delta j_1 (\beta _{11}\frac{r_0}{R_0})$ and the frequency of spherical oscillation correct upto second order in $\Delta$ for relativistic $\&$ non-relativistic case respectively can be written as,
\begin{equation}
\Omega_{sph}(rel)  \sim \omega_p\left[1  - \frac{3}{16}\frac{v_\phi ^2\Delta ^2}{c^2}j_1 ^2\left(\beta _{11}\frac{r_0}{R_0}\right) + \frac{v_\phi ^2\Delta ^2}{12\omega _p^2r_0^2}j_1 ^2\left(\beta _{11}\frac{r_0}{R_0}\right) \right] \label{eq:24}
\end{equation}
and
\begin{equation}
\Omega_{sph}(non-rel)  \sim \omega_p\left[1  + \frac{v_\phi ^2\Delta ^2}{12\omega _p^2r_0^2}j_1 ^2\left(\beta _{11}\frac{r_0}{R_0}\right) \right] \label{eq:25}
\end{equation}
Again following Dawson's argument \cite{1}, one can easily arrive at the same scaling law given by Eq.(\ref{eq:21}) for both relativistic and nonrelativistic case. The variation of phase mixing time $(\tau_{mix})$ as a function of applied amplitude $\Delta$ for relativistic and nonrelativistic spherical oscillations are shown in Fig-(\ref{fig:fig7}). In the figure numerical results have been displayed by points and the solid lines represent the scaling given by Eq.(\ref{eq:21}).

In all these cases we observe that, the analytical scaling law given by Eq.(\ref{eq:21}) shows a very good fit to the observed numerical results, thus vindicating our weakly relativistic calculations. 

\section{Summary} \label{sec:4}
\begin{figure}[htbp]
\center
\includegraphics[width=0.8\textwidth]{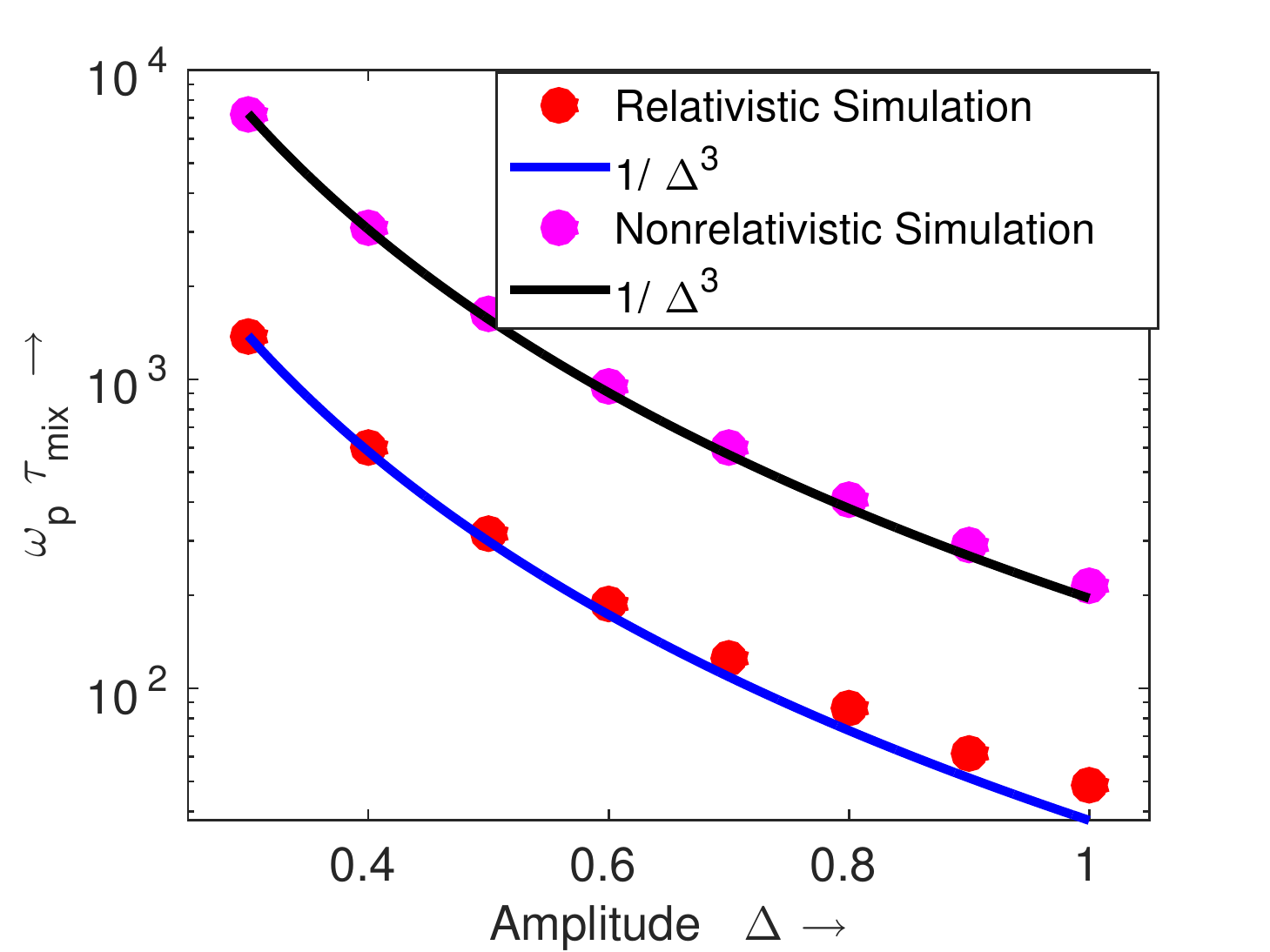}  
\caption{(Color online) Variation of phase mixing time of relativistic and nonrelativistic cylindrical oscillations as a function of amplitude of applied perturbation $\Delta$ [given by Eq.(\ref{eq:21})].}  
\label{fig:fig3} 
\end{figure}

\begin{figure}[htbp]
\center
\includegraphics[width=0.6\textwidth]{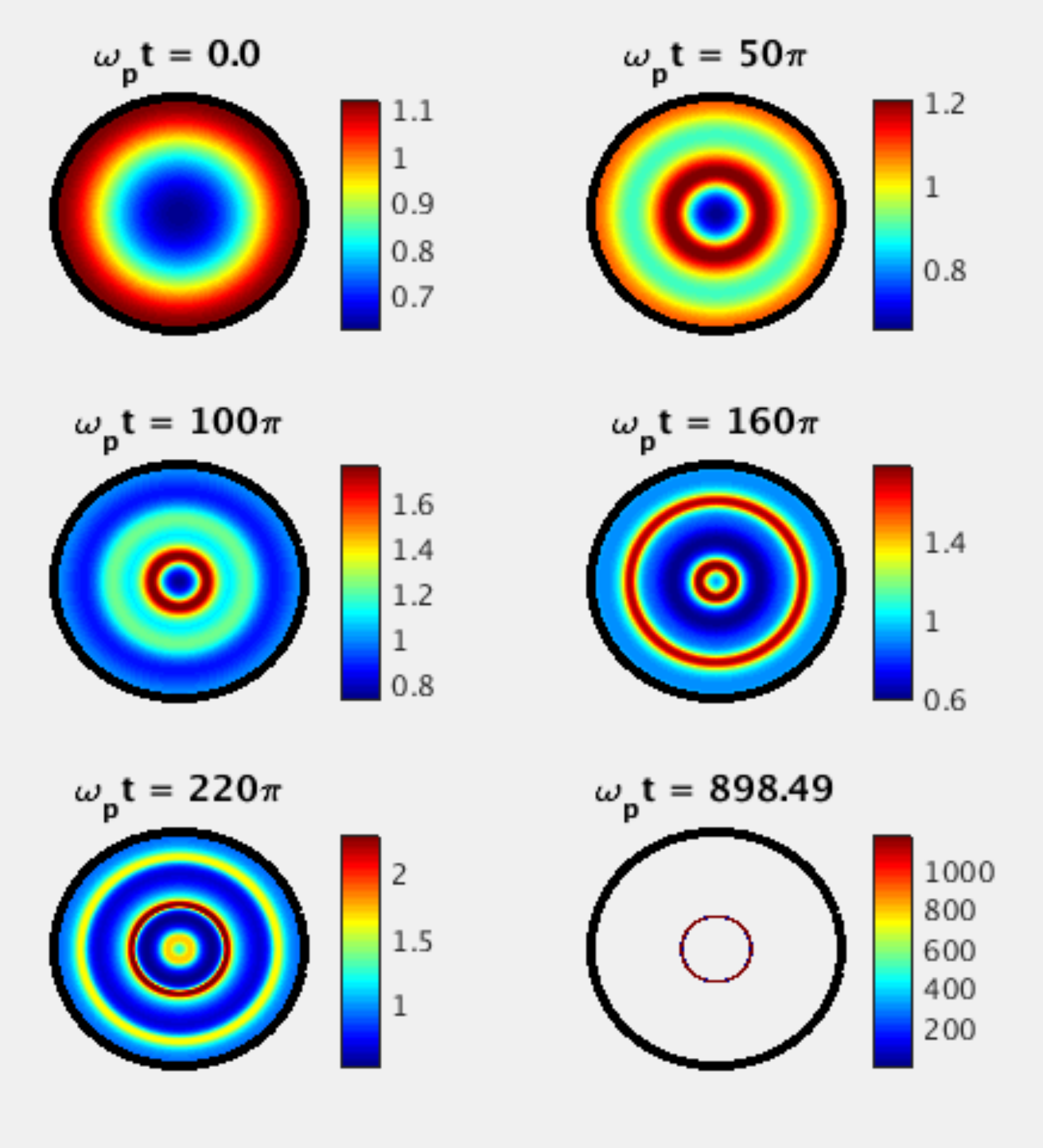}  
\caption{(Color online) Time evolution of density of relativistic spherical oscillation for amplitude $\Delta = 0.5$.}  
\label{fig:fig5} 
\end{figure}

\begin{figure}[htbp]
\center
\includegraphics[width=0.6\textwidth]{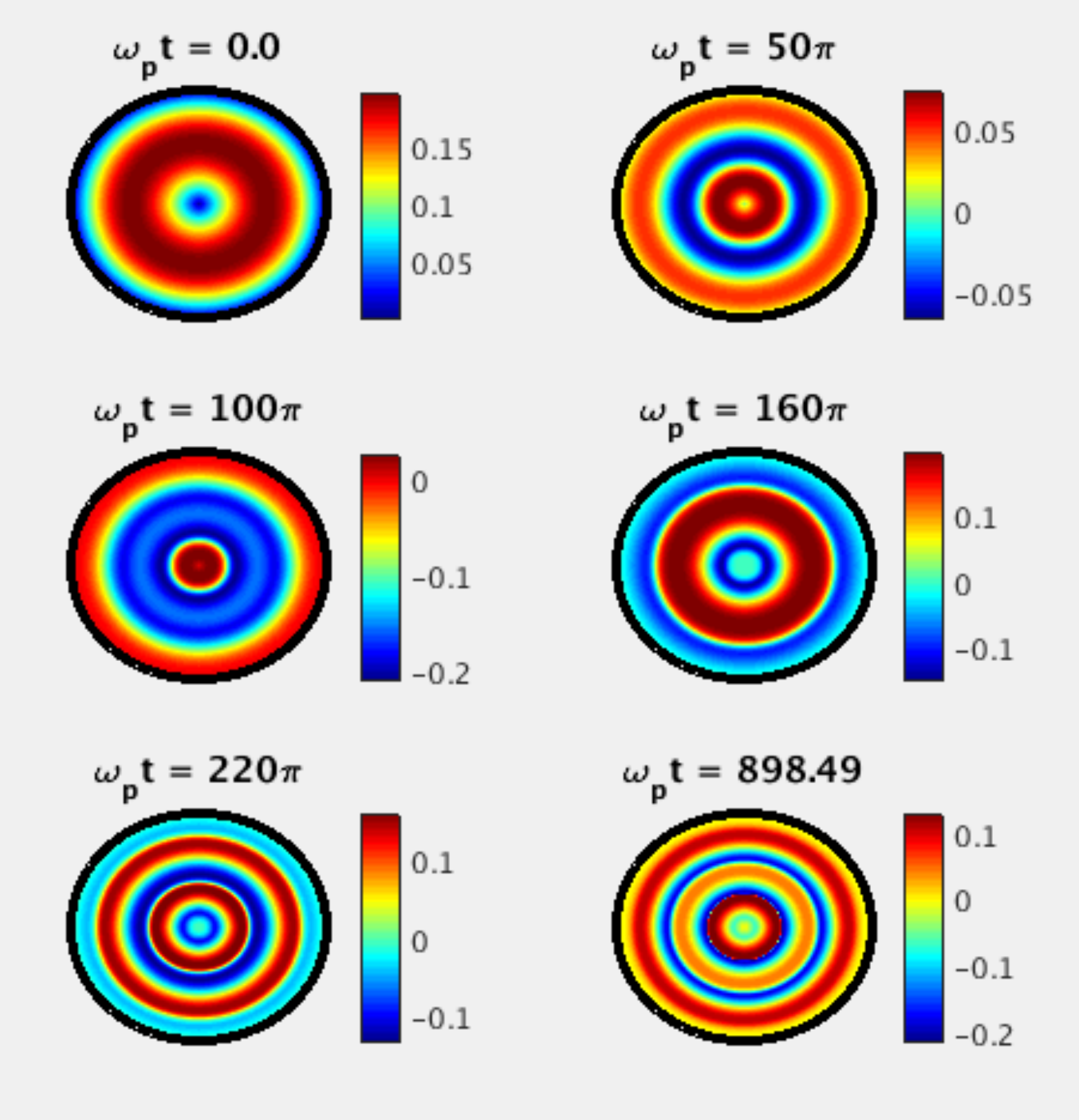}  
\caption{(Color online) Time evolution of electric field of relativistic spherical oscillation for amplitude $\Delta = 0.5$.}  
\label{fig:fig6} 
\end{figure}

\begin{figure}[htbp]
\center
\includegraphics[width=0.8\textwidth]{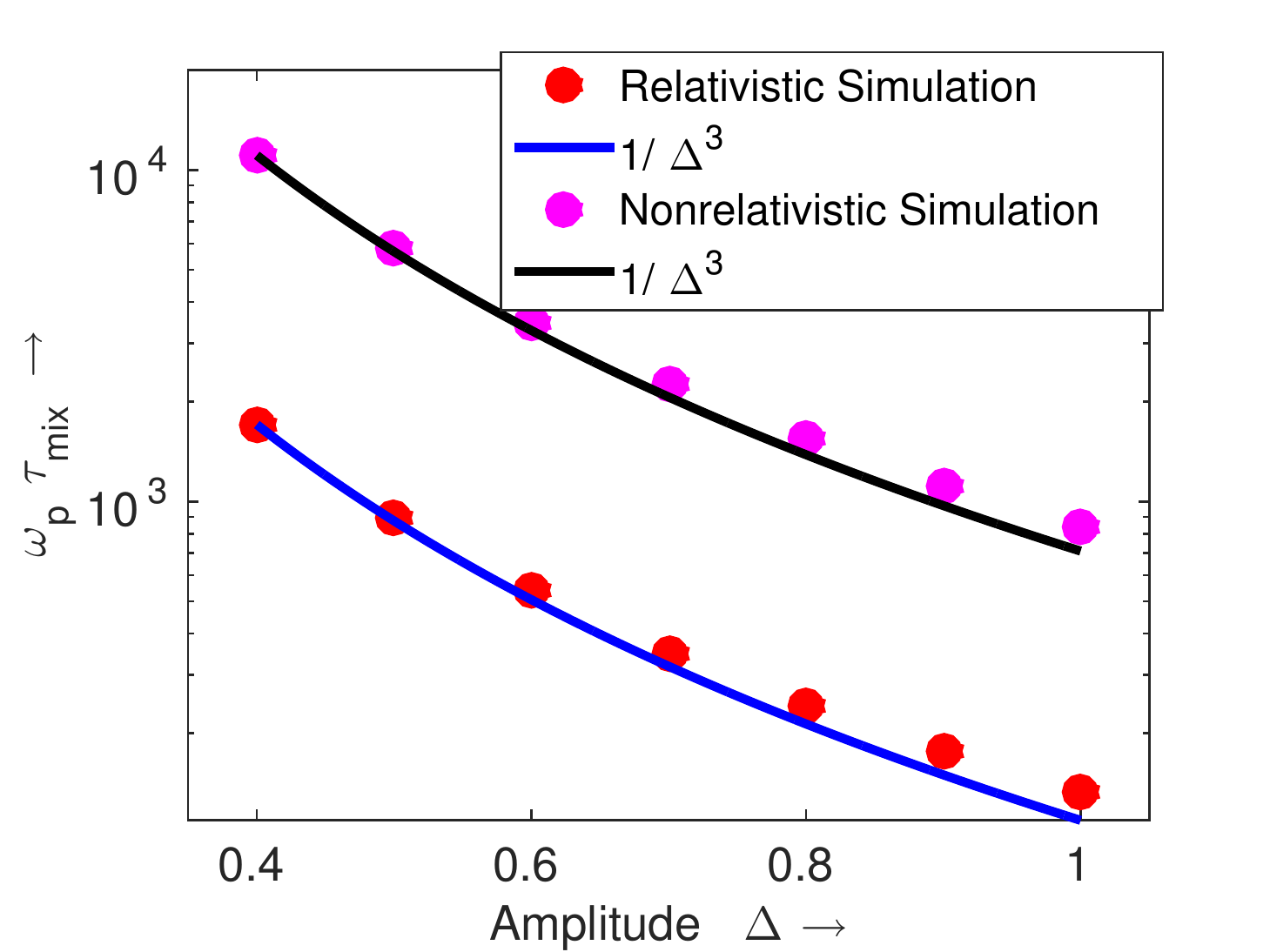}  
\caption{(Color online) Variation of phase mixing time of relativistic and nonrelativistic spherical oscillations as a function of amplitude of applied perturbation $\Delta$ [given by Eq.(\ref{eq:21})].}  
\label{fig:fig7} 
\end{figure}

 In this paper, we have demonstrated analytically and numerically the behaviour
of relativistically intense cylindrical and spherical plasma oscillations
using Dawson sheet model. Initial conditions are taken in terms of Bessel functions and Spherical Bessel functions to excite cylindrical and spherical oscillations respectively. This is because, as Bessel functions and Spherical Bessel functions are respectively the normal modes in cylindrical and spherical coordinate systems, thus any arbitrary perturbation in these systems can be written as a superposition of these basis functions in their respective coordinate systems. The expressions for frequencies have been given for both the cases and is found to be an explicit function of the equilibrium positions of the electron sheets. By performing numerical simulations it has been shown that the electron number density associated with cylindrical and spherical plasma oscillations grows sharply with time and after few plasma periods the density shows explosive behaviour due to the crossing of neighbouring electron sheets which is a signature of wave breaking. Analytical expression for the phase mixing time scale has been derived and it is observed that for both cases (cylindrical and spherical) phase mixing time scales with the cube of the oscillation amplitude, which indicates that in general scaling of phase mixing time with amplitude is independent of geometry of oscillation. Further we have found that inclusion of relativistic effects does not change the scaling of phase mixing time with amplitude of perturbation; it only hastens the process as compared to the non-relativistic case as depicted in Figs.(\ref{fig:fig3}) and (\ref{fig:fig7}). Thus cylindrical and spherical oscillations initiated by an arbitrary density perturbation (or electric field perturbation) will always phase mix and the phase mixing time scale can be estimated from the scaling law given by Eq.(\ref{eq:21}).

%%%%%%%%%%%%%%%%%%%%%%%%%%%%%%%%%%%%%%%%%%%%%%%%%%%%%%%%%
%%%%%%%%%%%%%%%%%%%%%%%%%%%%%%%%%%%%%%%%%%%%%%%%%%%%%%%%%

%%%%%%%%%% FIGURES %%%%%%%%%%%
\newpage
\vspace*{\fill}
\vspace*{\fill}

\newpage
\vspace*{\fill}
\vspace*{\fill}

\newpage
\vspace*{\fill}
\vspace*{\fill}

%\newpage
%\vspace*{\fill}
%\begin{figure}[htbp]
%\center
%\includegraphics[width=0.8\textwidth]{fig_4}  
%\caption{(Color online) Variation of phase mixing time of nonrelativistic cylindrical oscillation as a function of amplitude of applied perturbation $\Delta$[Eq.\ref{eq:21}].}  
%\label{fig:fig4} 
%\end{figure}
%\vspace*{\fill}

\newpage
\vspace*{\fill}
\vspace*{\fill}

\newpage
\vspace*{\fill}
\vspace*{\fill}

\newpage
\vspace*{\fill}
\vspace*{\fill}

%\newpage
%\vspace*{\fill}
%\begin{figure}[htbp]
%\center
%\includegraphics[width=0.8\textwidth]{fig_8}  
%\caption{(Color online) Variation of phase mixing time of nonrelativistic spherical oscillation as a function of amplitude of applied perturbation $\Delta$[Eq.\ref{eq:21}].}  
%\label{fig:fig8} 
%\end{figure}
%\vspace*{\fill}

%%%%%%%%%%%%%%%%%%%%%%%%%%%%%%%%%%%%%%%%%%%%%%%%%%%%%%%%%%%%
%%%%%%%%%%%%%%%%%%%%%%%%%%%%%%%%%%%%%%%%%%%%%%%%%%%%%%%%%%%%

\begin{thebibliography}{1}
\bibitem{a}
R.C. Davidson, \textit{Methods in Nonlinear Plasma Theory} (Academic, New York, 1972)
\bibitem{b}
P. Gibbon, \textit{Short Pulse Laser Interaction with Matter} (Imperial College Press).
\bibitem{7}
A.I. Akhiezer and R.V. Polovin, Sov. Phys. JETP {\bf{3}}, 696 (1956).
\bibitem{e}
A. Modena, Z. Najmudin, A.E. Dangor, C.E. Clayton, K.A. Marsh, C. Joshi, V. Malka, C.B. Darrow, C. Danson, D. Neely and F.N. Walsh, Nature {\bf{377}}, 606 (1995).
\bibitem{f}
V. Malka, S. Fritzler, E. Lefebre, M.M. Aleonard, F. Burgy, J.P. Chambaret, J.F. Chemin, K. Krushelnick, G. Malka, S.P.D. Mangles, Z. Najmudin, M. Pittman, J.P. Rousseau, J.N. Scheurer, B. Walton and A.E. Dangor, Science {\bf{298}}, 1596 (2002).
\bibitem{g}
J. Faure, C. Rechatin A. Norlin, A. Lifschitz, Y. Glinec and V. Malka, Nature(London) {\bf{444}}, 737 (2006).
\bibitem{i}
C. Rechatin, J. Faure, A. Ben-Ismail, J. Lim, R. Fitour, A. Speckam, H. Videau, A. Tafzi, F. Burgy and V. Malka, Phys. Rev. Lett. {\bf{102}}, 164801 (2009).
\bibitem{j}
E. Esarey, C.B. Schroeder and W.P. Leemans, Rev. Mod. Phys. {\bf{81}}, 1229 (2009).
\bibitem{c}
T. Tajima and J.M. Dawson, Phys. Rev. Lett. {\bf{43}}, 267 (1979).
\bibitem{a1}
Max Tabak, James Hammer, Michael E. Glinsky, William L. Kruer, Scott C. Wilks, John Woodworth, E. Michael Campbell, Michael D. Perry and Rodney J. Mason, Phys. Plasmas {\bf{1}}, 1626 (1994).
\bibitem{a2}
R. Kodama, P. A. Norreys, K. Mima, A. E. Dangor, R. G. Evans, H. Fujita, Y. Kitagawa, K. Krushelnick, T. Miyakoshi, N. Miyanaga, T. Norimatsu, S. J. Rose, T. Shozaki, K. Shigemori, A. Sunahara1, M. Tampo, K. A. Tanaka,5, Y. Toyama, T. Yamanaka and M. Zepf, Nature {\bf{412}}, 798 (2001).
\bibitem{ssnf}
S. Sengupta, A. S. Sandhu, G. R. Kumar, A. Das and P. K. Kaw, Nucl. Fusion {\bf{45}}, 1377 (2005).
\bibitem{1}
J.M. Dawson, Phys. Rev.{\bf{113}}, 383 (1959).
\bibitem{4}
S. Sengupta, V. Saxena, P.K. Kaw, A. Sen and A. Das, Phys. Rev. E {\bf{79}}, 026404 (2009).
\bibitem{5}
S. Sengupta, P. Kaw, V. Saxena, A. Sen and A. Das, Plasma Phys. Controlled Fusion {\bf{53}}, 07414 (2011).
\bibitem{6}
Sudip Sengupta, AIP Conf. Proc. 1582, 191–200 (2014)
\bibitem{8}
Prabal Singh Verma, Sudip Sengupta and Predhiman Kaw, Phys. Rev. Lett. {\bf{108}}, 125005 (2012).
\bibitem{k}
Chandan Maity, Anwesa Sarkar, Padma Kant Shukla and Nikhil Chakrabarti,
 Phys. Rev. Lett. {\bf{110}}, 215002 (2013).
\bibitem{16}
Arghya Mukherjee and Sudip Sengupta , Phys. Plasmas. {\bf{21}}, 112104 (2014).
%%%%%%%%%%%%%%%%%%%%%%%%%%%%%%%%%%%%%%%%%%%%%%%%%%%%%%%%%%%%%%%%%%%%%%%%%%%
%%%%%%%%%%%%%%%%%%%%%%%%%%%%%%%%%%%%%%%%%%%%%%%%%%%%%%%%%%%%%%%%%%%%%%%%%%%
%%%%%%%%%%%%%%%%%%%%%%%%%%%%%%%%%%%%%%%%%%%%%%%%%%%%%%%%%%%%%%%%%%%%%%%%%%%
\bibitem{3}
E. Infeld and G. Rowlands, Phys. Rev. Lett. {\bf{62}}, 1122 (1989).
\bibitem{arsu}
Arghya Mukherjee and Sudip Sengupta , Phys. Plasmas. {\bf{23}}, 092112 (2016).
\bibitem{ref244}
E. Infeld, G. Rowlands, Phys. Rev. Lett. {\bf{62}}, 1122 (1989).
\bibitem{ref233}
Sudip Sengupta and Predhiman K. Kaw, Phys. Rev. Lett. {\bf{82}}, 1867 (1999).
\bibitem{drake}
 J. F. Drake, Y. C. Lee, K. Nishikawa, N. L. Tsintsadze, Phys. Rev. Lett. {\bf{36}}, 196 (1976).
\bibitem{17}
 E. V. Chizonkov, A. A. Frolov nad L. M. Gorbunov, Rus. J. Numer. Anal. Math. Model {\bf{23}}, 455 (2008).
\bibitem{18}
L.M. Gorbunov, A.A. Frolov, E.V. Chizkonov and N.E. Andreev, Plasma Phys. Rep. {\bf{36}}, 345 (2010).
\bibitem{19}
S. V. Bulanov, A. Maksimchuk, C. B. Schroeder, A. G. Zhidkov, E. Esarey, Phys. Plasmas. {\bf{19}}, 020702 (2012)
\bibitem{20}
Sergei V. Bulanov, Timur Zh. Esirkepov, Masaki Kando, James K. Koga, Tomonao Hosokai, Alexei G. Zhidkov and Ryosuke Kodama, Phys. Plasmas. {\bf{20}},083113 (2013).
\bibitem{21}
J. R. Marquès, J. P. Geindre, F. Amiranoff, P. Audebert, J. C. Gauthier, A. Antonetti, and G. Grillon, Phys. Rev. Lett. {\bf{76}}, 3566 (1996).
\bibitem{22}
J. R. Marquès, F. Dorchies, P. Audebert, J. P. Geindre, F. Amiranoff, J. C. Gauthier, G. Hammoniaux, A. Antonetti, P. Chessa, P. Mora, and T. M. Antonsen, Jr., Phys. Rev. Lett. {\bf{78}}, 3463 (1997).
\bibitem{23}
Ali Hasan Nayfeh, \textit{Perturbation Methods} (Wiley and Sons, New York)
\bibitem{25}
J.M. Dawson, Phys. Fluids. {\bf{5}}, 445 (1962).
%\bibitem{24}
%R. W. C. Davidson and P. P. J. M. Schram, Nucl. Fusion {\bf{8}}, 183 (1968)
\bibitem{14}
 George B. Arfken, Hans J. Weber and Frank E. Harris , \textit{Mathematical Methods for Physicists, 4th ed.} (Academic, San Diego, 1995)
\bibitem{15}
 J. D. Jackson , \textit{Classical Electrodynamics, 3rd ed.} (Wiley and Sons, New York, 1998)
\bibitem{10}
A. W. Travelpiece and R. W. Gould, Journal of Appl. Phys. {\bf{30}}, 1784 (1959)
\bibitem{11}
Wallace M. Manheimer, Phys. Fluids. {\bf{12}}, 2426 (1969)
\bibitem{12}
J. Juul Rasmussen, Plasma. Phys. {\bf{20}}, 997 (1978)
\bibitem{13}
Thomas P. Hughes and Edward Ott, Phys. Fluids. {\bf{23}}, 2265 (1980)


\end{thebibliography}
\end{document}